Origins of Bipedalism

Kwang Hyun Ko

Hanyang University Research

Author's note

Correspondence concerning this article should be addressed to Hanyang University

Tel: (82) 010-8517-1288 | Email: highwaytolife2@gmail.com /kwhyunko@gmail.com/

highway2@hanyang.ac.kr

Abstract

Abstract: The following manuscript reviews various theories of bipedalism and provides a holistic answer to human evolution. There are two questions regarding bipedalism: i) why were the earliest hominins partially bipedal? and ii) why did hominins become increasingly bipedal over time and replace their less bipedal ancestors? To answer these questions, the prominent theories in the field, such as the savanna-based theory, the postural feeding hypotheses, and the provisioning model, are collectively examined. Because biological evolution is an example of trial and error and not a simple causation, there may be multiple answers to the evolution of bipedalism. The postural feeding hypothesis (reaching for food/balancing) provides an explanation for the partial bipedalism of the earliest hominins. The savannah-based theory describes how the largely bipedal hominins that started to settle on the ground became increasingly bipedal. The provisioning model (food-gathering/monogamy) explains questions arising after the postural feeding hypothesis and before the savannah theory in an evolutionary timeline. Indeed, there are no straight lines between the theories, and multiple forces could have pushed the evolution of bipedalism at different points. Finally, this manuscript states that the arboreal hominins that possessed ambiguous traits of bipedalism were eliminated through choice and selection. Using the biological analogy of the okapi and giraffe, I explain how one of the branches (Homo) became increasingly bipedal while the other (Pan) adapted to locomotion for forest life by narrowing the anatomical/biological focus in evolution.

Origins of Bipedalism

## Introduction

Bipedalism is an essential adaptation of the *Hominin* progeny that is considered the major force behind several skeletal changes shared by all bipedal hominins (Lovejoy 1988). There are different hypotheses that explain how and why bipedalism evolved in humans. Similarly, the timing of the evolution of bipedalism is debated. The possible reasons for the evolution of human bipedalism include the freeing of the hands to use and carry tools, threat displays, sexual dimorphism in food gathering, and changes in climate and habitat (from jungle to savanna).

## Bipedalism: New Perspective

As in other species, several characteristics of the ape-like hominin ancestors were advantageous for their survival. Human bipedalism was driven by the simple Darwinian principle of natural selection. Hominins did not consciously become bipedal for a specific reason. Instead, eons of time allowed the evolution of bipedalism in humans because it was a favorable trait (Auletta et al., 2011). Specifically, a distinctive set of observable traits in each species constitute characteristics that have lasted through natural selection out of the countless mutation traits that were observed during the timeline of the species (Ayala, 2007). *Figure 1*

This issue becomes more complex when we attempt to investigate the actual process of 'Natural Selection' or 'Darwinism' because several factors intervene during millions of years (Darwin, 1963). Such factors may be interaction with animals, the avoidance of competition and/or the effective protection of offspring. Natural selection may also be influenced by changes in the environmental settings (Miller, 1995). 'Natural selection' is, unfortunately, a vague term that includes several factors. However, because of the wide definition of Darwinism, it does not have to be a single discrete principle or hypothesis that provides holistic explanations of how

'survival' occurs.

Scholars tend to fixate on causation, where A caused B (in this case, B is bipedalism), and are currently attempting to find an exact answer for the cause of bipedalism. However, the principles of evolution do not necessarily operate like cause and effect. Biological evolution is an example of trial and error (Wright, 1932), i.e., if a trait works, it remains. Again, various traits were observed during the evolutionary timeline, but only a few remained.

The reason why bipedalism remained is because it was beneficial for the efficient survival of both the unit itself and its offspring. There are several theories debating human bipedalism (Tuttle, 2015). However, if the evidence explains how bipedalism helped our human ancestors survive, it is sensible to believe that there may be multiple answers to the question of the evolution of bipedalism. Specifically, the incremental change of bipedalism could have aided the actual survival of the animal unit through its adaption to new environments, the avoidance of predators, the conservation/gaining of more nutriments, and the successful protection of the progeny by the parental unit. It is possible that bipedalism provided a variety of benefits to the hominin species.

Furthermore, there are several paths through which evolution could have benefited the survival of our ancestors. However, the hominins are the only bipedal species out of all of the great apes (Harcourt-Smith, 2007). Therefore, it is important to note that this change was advantageous for humans but not advantageous for the other great apes.

Finally, a retracing of the evolutionary traits backwards in an evolutionary timeline from the modern *Homo sapiens* to the Neanderthals, the *Homo erectus*, the *Australopithecus* and, finally, the *Ardipithecus* revealed that the latter species bears a closer resemblance to the last common ancestor of chimpanzees and humans. The unsettled dispute lies with the earliest bipedal hominin, which could be Sahelanthropus, Ardipithecus or Orrorin (Su, 2013). In addition, several studies have indicated that the extinct hominin Ardipithecus, which was extremely similar to the common chimpanzee ancestor, possessed the ability to walk on two feet while spending time in the trees. Indeed, the first hominin or the first common ancestor was partially bipedal, i.e., it possessed a limited ability to walk on two feet. *Figure 2*

The following paragraphs will review several prominent theories of bipedalism. The different models of bipedalism will be examined in accordance with the factors of natural selection. Moreover, a comprehensive approach based on an evolutionary timeline of other great apes in the Hominini tribe and even in the Homininae subfamily will be explained considering the aforementioned perspective of multiple answers.

**The savanna-based theory**

The savanna-based theory was one of the earliest models to explain the origins of bipedalism and gathered support from several anthropologists (Dart, 1925). It mainly suggests that the early hominids were forced to adapt to an open savanna after they left the trees by walking erect on two feet (Shreeve, 1996). According to this theory, the evolution of bipedal locomotion would have been helpful in a savanna because the posture would allow hominins to watch over tall grasses, hunt effectively or be aware of predators. Unfortunately, the fossil record shows that the early bipedal hominines were still adapted to climbing trees, and research has indicated that bipedalism evolved in trees.

**The postural feeding hypotheses**

The second model is the postural feeding hypothesis, which was recently proposed by Kevin Hunt at Indiana University. He asserts that bipedal movements may have evolved into regular habits because they were convenient for obtaining food and keeping balance (Hunt, 1994). It has been observed that chimps are only bipedal when they eat. Chimpanzees would reach up for fruit hanging from trees, and orangutans used their hands to stabilize themselves while navigating thinner branches (Stanford, 2006). Hunt asserts that *Australopithecus afarensis* has hand and shoulder features that demonstrate hanging habits, whereas their hip and hind limb clearly indicate bipedalism. Because a bipedal posture was utilized for grabbing from an overhead branch and harvesting food, Hunt argues that bipedalism evolved more as a feeding posture than as a walking posture.

## Review: Introducing a new perspective

If we were to retrace the steps back to our common ancestor, we could find clues to address these perplexing issues and theories. The earliest hominins, *Sahelanthropus* and *Ardipithecus*, have been suggested to have been bipedal and partly arboreal (Nelson, 2013). Interestingly, extinct hominins that were close to the common chimpanzee ancestor were partially bipedal. The evolutionary momentum gradually pushed the common ancestor, which was limitedly bipedal and arboreal, to become chimps that were mostly arboreal with limited bipedal motion (quadrupedal mostly on ground) in one branch and hominins that were mostly terrestrial with full bipedal locomotion in the other evolutionary branch. Evolution, therefore, did not have a single direction from the common ancestor toward *Homo sapiens*. Ancestors of both chimps and humans that apparently possessed ambiguous traits of humans and chimps evolved in two ways: one toward chimpanzees, which include great chimpanzees and bonobos, and the other toward *Homo sapiens* (Patterson et al., 2006).

Importantly, hominins slowly evolved to walk like modern humans over a continuous scale. Therefore, the important question is not why the earliest hominins were partially bipedal but rather why hominins became more bipedal over time and replaced their less-bipedal ancestors. This specific evolutionary trait of bipedalism was not necessary for chimps and their extinct ancestors that lived on trees.

Despite the alleged lack of evidence, the fact remains that full bipedalism has not been documented in other great apes. Other Hominidae/great apes species, such as gorillas, orangutans, and chimps, but not humans spend much time in trees. Chimps, for example, are agile climbers and nest in trees to rest around noon and sleep at night. During the day, gorillas climb trees, swing from branches, and chase one another. Most arboreal great apes, such as orangutans, spend nearly all of their time in trees. Bipedalism (full bipedalism observed in *Homo erectus* and modern humans) is not a beneficial trait when moving from one tree to another in an arboreal life.

In addition, a change in environment (moving away from trees) cannot be a cause of the partial bipedalism in early hominins, as suggested by fossil evidence. Nevertheless, when hominins started to settle on the ground, the savannah-based theory can provide an explanation for why hominins evolved to walk like modern humans, replacing their less-bipedal ancestors (Dart, 1925). Furthermore, the savannah-based theory incorporates several models of bipedalism, such as the sentinel response, threat display, and endurance running, all of which provide general evidence for how bipedalism aided the survival of hominins in the savannah.

Another model is the postural feeding hypothesis, which is supported by evidence from several studies. However, there are logical problems associated with it. There are possibly two aspects of bipedalism: i) how were the earliest hominins partially bipedal in the first place? and ii) why did the hominins become increasingly bipedal over time?

Hunt's theory cannot explain the second aspect of why partially bipedal hominins evolved to walk like modern humans. He stated the advantage of obtaining food from branches or balancing in an arboreal task (Hunt, 1992). However, hominins evolved to walk like modern humans on the ground, not in trees. The evolutionary change that was driven by balancing and reaching in trees should not have affected hominins that were abandoning the arboreal life. Hominins continuously evolved to possess terrestrial adaptations and eventually lived on the ground.

However, Hunt's theory can explain the origin of bipedalism, i.e., how the last common ancestor of chimpanzees and humans, or why the ancient Miocene hominins were partially bipedal. This theory can explain that some Hominidae became partially bipedal for the benefit of balancing or obtaining food from branches.

**The threat model**

The original proponents of this model theorized that bipedalism originated as a natural defense strategy for early hominids. Hominids were attempting to stay as visible as possible according to instincts of aposematism or warning displays and intimidation of potential predators (Jablonski & Chaplin, 1993). Several morphological and behavioral developments were undertaken to exaggerate visual signals, such as the upright bipedal posture, longer legs, and synchronous body movements.

**The thermoregulatory model**

Peter Wheeler proposed the thermoregulatory model, a model that states that bipedalism would increase the amount of body surface area, which helps dissipate heat and reduces heat gain (Wheeler, 1984). Hominins gain access to more favorable wind speeds and temperatures by being higher above the ground. In addition to reducing the body surface exposed to heat, greater wind flows result in a higher temperature loss, which makes the organism more comfortable.

**Review**

        Although bipedalism appears to be a favorable trait, this locomotion in hominins has offered certain drawbacks to survival. The maximum sprint speed of fully bipedal humans is strikingly slower than that of many animals (Powell, 2007). Bipedalism was seemingly advantageous, but the surprisingly slow speed observed is attributed to the human anatomy that allowed bipedalism. The average human speed is slower than that of other apes. Such slow speed was a dangerous trait for survival because hominins became vulnerable to carnivores. Full bipedalism was eventually advantageous, but at the beginning, bipedalism exposed humans to the risk of predation. An effective bipedalism theory should state the advantages and explain how a trait that offered an advantage outran potential disadvantages associated with survival.

        Other theories that are not mentioned here, in addition to warning display and heat loss, provide a broad reasoning that can be applied not only for hominins and primates but for any animal species. The theory should state not only possible advantages of bipedalism but also why a particular trait would have been selected in hominins over millions of years.

        In addition to primates, several species, such as frogs, snakes, the northern white-faced owl, Phasmatodea, and praying mantis, perform warning displays (Eisner & Grant, 1981). However, this does not explain why only humans would walk on two feet to adapt to a warning display. Similar issues remain to be resolved to explain the thermoregulatory model. Thermoregulation (storing fat, panting, estivation, hibernation, etc.) is an important aspect of survival in many species, but not all animals have become bipedal to lose heat or to control their body temperature. The two hypotheses do not provide a clear reasoning for why humans are the only species to have achieved bipedalism.

        The foregoing theories and other unmentioned theories, such as heat loss, warning,

sentinel behavior or running endurance, automatically presume that humans were already adapted to a terrestrial (savannah) life, whereas the earliest hominins were largely tree-dwellers. Several of these theories, therefore, would be incorporated into the savannah-based theory. A few studies have demonstrated that an intimidating visual display and heat loss were possible advantages of bipedalism, but the logic would not remain in an opposite direction.

**Overcoming the disadvantage**

The clear advantage of bipedalism was the possibility for ancient hominin species to use their hands. With the evolution of bipedalism, this special advantage was evolved only by primates. This advantage offered a benefit that overcame the fatal disadvantage of slow speed. Several species could not have utilized their hands for effective provisioning or tool use, even if they had become bipedal through evolution. However, the prehensile hands and feet of primates evolved from the mobile hands of the semi-arboreal tree shrews that lived approximately 100 million years ago and enabled provisioning in ape-like ancestors (Schmidt & Lanz, 2004). Similar to humans, modern-day chimpanzees have a limited ability to use their limbs and even sticks to obtain termites in a manner similar to human fishing.

**Provisioning model**

The last prominent bipedalism theory, which was proposed by Owen Lovejoy, is the provisioning model. Lovejoy, the director of the Matthew Ferrini Institute, suggested a modified version of Darwin's explanation. What would have been so advantageous about using two hands? Lovejoy proposed that walking on two legs was a main adaptation for pair-bonding to succeed because carrying with two hands was effective for food transport (Lovejoy, 1988).

Lovejoy's theory also proposed that sexual dimorphism suggests that food gathering would improve the infant survival rate. Males were responsible for provisioning females, whereas females protected their offspring (Lovejoy, 1980). Females would mate exclusively with the provisioning male, and other males would no longer need to fight with each other over females. Therefore, the males' jagged, blade-like canine teeth diminished over time. Several studies have demonstrated that chimpanzees can carry twice as many nuts in a bipedal position than when walking on all fours (Carvalho, et al. 2012). Anthropological evidence also supports this theory. The downsizing of male canine teeth, the decrease in antagonistic behavior and the body size dimorphism corroborate Lovejoy's theory.

## Review and Conclusion

As mentioned previously, there may be multiple answers to bipedalism, and there are two aspects of bipedal evolution: (1) the fact that ancient hominins were already partially bipedal and (2) the fact that hominins evolved full bipedalism. Although the postural feeding theory provides an explanation for the first aspect, the savanna-based theory can provide an answer to why hominins became increasingly bipedal over time. Lovejoy's provisioning model lies between these two theories. Hunt's theory, which suggests that bipedalism involved reaching for food and balancing on trees, would logically fall before using hands for provision. The early hominins spent time in trees, but the species eventually evolved to walk like modern humans on the ground. The evolutionary momentum that was driven by balancing and reaching in trees should have affected the early hominins. The provisioning model demonstrates how hominins became more bipedal over time not only by food gathering but also by provisioning infants (monogamy). Nonetheless, the provisioning model does not have sufficient evidence explaining why hominins would have begun to walk like modern humans and have lost all adaptations to arboreal life. To this gap is where the savannah-based theory contributes its explanations. When largely bipedal hominins started to settle on the ground, the savannah based-theory would be the explanation for their full bipedalism. The savannah based-theory includes various other models that already assume that hominins started to live a terrestrial life, such as sentinel behavior, threat, running endurance and thermoregulatory models. The general order of the theories is the following: postural-feeding, provisioning, and savannah-based theories. However, there are no straight lines between these theories, and it is possible that the three forces worked together at one point.

**Addition to the Conclusion: Biological Analogy of Bipedal Evolution**

First, there were arboreal hominids that possessed ambiguous traits of bipedalism. These were gradually replaced by two lines of species: one consisting of *Pan* species and the other comprising hominins. Hominin-like species and modern chimp-like species gradually evolved to undergo specific adaptations to live on the ground and trees, and the new hominins presented a survival advantage over their common ancestors. The ambiguous traits were eliminated through choice and selection. By narrowing its anatomical/biological focus over evolutionary time, one branch became more bipedal, while the other adapted to a quadrupedal locomotion.

However, when the split between the two species became clear, the hominins and chimpanzees would not have competed for resources.

The stated biological relationship between chimps and humans is similar to the remarkable relationship between the okapi and the giraffe. Similar to the unique adaptation of bipedal locomotion that was only observed in *Homo* species, the giraffe's long neck is also an evolutionary product exclusive to this species (Badlangana et al., 2007). The okapi and the giraffe are currently the only living members of the Giraffidae family. Although the short-necked okapi's outer appearance resembles a zebra, the okapi is the closest surviving species to the giraffe. Apparently, Darwinian natural selection has led the ancestral giraffes with long necks to reproduce and pass on their genes because they had a competitive advantage that enabled them to reach higher branches. Consequently, the giraffe ancestors fed on acacia leaves and spread through the savannah where tall trees grow. In contrast, the long-neck adaptations became futile for the okapi species, which, ultimately, inhabited canopy forests and fed on buds, grasses, ferns, fruits, or fungi (Hart, J. & Hart, T., 1989). *Figure 3*

Dear Editors,

The author is currently receiving treatment at a hospital. A rapid decision to start the review process for this paper would be greatly appreciated.

Similarly, I would like to acknowledge any comments, feedback or suggestions that the editor or reviewer may have, even if the editors decide to reject this article.

Sincerely,